# Indium tin oxide nanoparticle:TiO$_2$:air layers for one-dimensional multilayer photonic structures

Ilka Kriegel [1], Francesco Scotognella [2,3,*]

[1]*Department of Nanochemistry, Istituto Italiano di Tecnologia (IIT), via Morego, 30, 16163 Genova, Italy*
[2]*Dipartimento di Fisica, Politecnico di Milano, Piazza Leonardo da Vinci 32, 20133 Milano, Italy*
[3]*Center for Nano Science and Technology@PoliMi, Istituto Italiano di Tecnologia, Via Giovanni Pascoli, 70/3, 20133, Milan, Italy*
*email address: francesco.scotognella@polimi.it

**Abstract**
In this work we study the optical properties, by means of the transfer matrix method, of one-dimensional photonic crystals in which layers of silica nanoparticles are alternated with layers of indium tin oxide nanoparticle (ITO) / titania nanoparticle mixture. The dielectric function of the mixed ITO/TiO$_2$ nanoparticle layer is carefully accounted for with a generalized Rayleigh equation for the ternary mixture ITO:TiO$_2$:air. We have studied the light transmission of the multilayer photonic crystal as a function of the ITO/TiO$_2$ ratio. We observe that, by increasing the ITO content in the three-phase mixture, the intensity of the plasmon resonance in the near infrared (NIR) increases and the intensity of the photonic band gap (visible) decreases. Thus, our study is of major importance for the realization of electrochromic smart windows, in which separate and independent NIR and visible light control is required.

**Introduction**
Porous nanoparticle based multilayer photonic crystals combine very simple and versatile methods, as for example spin coating [1,2] and sputtering [3], with the arise of the photonic band gap [4–6] and with the possibility to exploit the nanoparticle features or the porosity of the device for sensing, lasing, and switching [7,8]. Multiple optical functionalities can be obtained by choosing the proper nanoparticles forming the thin films of the multilayer structure. In particular the refractive index contrast between the two layers accounts for the photonic stop band (in the visible spectral range) and can be adjusted to deliver a targeted optical response. This becomes particularly interesting when implementing refractive index tunable materials, such as transparent conducting oxides (TCOs), whose optical response can be manipulated by the application of an electric [9] or an electrochemical potential [10]. A very interesting property of such materials is their near infrared (NIR) response due to the free carrier density in the range of $10^{26} - 10^{27}$ m$^{-3}$, which results in plasmonic absorption. Carriers are injected in a capacitive manner through the applied electric (electrochemical) potential, adding to the predefined carrier density, and thus, resulting in enhanced NIR absorption [11]. In a porous multilayer photonic crystal, this modification of the dielectric response automatically also affects the refractive index contrast and therefore provides a means to actively manipulate the photonic stop band [12,13].
For the application of such solution processed thin film structures as electrochromic materials it is of major importance to gain the independent control over the NIR and visible spectral range, as to produce smart windows that can expel either heat (NIR) or sunlight (visible) separately by applying a potential. Recent works addressed this target by implementing NbOx glasses as matrix [14]. This involves an additional chemical step in the coloration, and thus, the devices become more prone to degradation. Implementing a physical effect in the coloration therefore is a highly attractive approach. In this work, we study the light transmission properties of a multilayer photonic crystal made by alternating silica nanoparticle layers of electrochemically tunable indium tin oxide (ITO) nanoparticle / titania nanoparticle layers. The implementation of the three material mixture (ITO:TiO2:air) delivers an additional ingredient that allows fine tuning of the optical response that is composed of the photonic band gap in unison with the plasmon resonance. Notably, we predict that the tuning of the ITO nanoparticle dielectric function, and thus the overall refractive index of the three-phase mixture layer, has influence both on the plasmon resonance of ITO and the photonic band gap. Remarkably, for the ITO:TiO2:air 60:10:30 ratio the increase of the number of carrier of ITO induce a simultaneous increase of the plasmon resonance intensity and decrease of the photonic band gap intensity.

**Methods**
*ITO:TiO2:air layer refractive index*: According to the Drude model to describe the plasmonic response [15], the frequency dependent complex dielectric function of ITO can be written as:

$$\varepsilon_{ITO,\omega} = \varepsilon_{1,\omega} + i\varepsilon_{2,\omega} \tag{1}$$

where

$$\varepsilon_{1,\omega} = \varepsilon_\infty - \frac{\omega_P^2}{(\omega^2 - \Gamma^2)} \tag{2}$$

and

$$\varepsilon_{2,\omega} = \frac{\omega_P^2 \Gamma}{\omega(\omega^2 - \Gamma^2)} \tag{3}$$

where $\omega_P$ is the plasma frequency and $\Gamma$ is the free carrier damping. The plasma frequency is $\omega_P = \sqrt{Ne^2/m^*\varepsilon_0}$, with $N$ number of charges, $e$ the electron charge, $m^*$ the effective mass and $\varepsilon_0$ the vacuum dielectric constant. For ITO, we use $N = 2.49 \times 10^{26}$ $charges/m^3$ (and we start from this value when we increase the number of carriers in this study), $m^* = 0.4/m_0$ $kg$. For the free carrier damping, we use $\Gamma = 0.1132$. The parameters for ITO nanoparticles are taken from [11,16]; the value of the high frequency dielectric constant, $\varepsilon_\infty = 4$, is taken from [17]. The wavelength dependent refractive index of TiO2 can be written [18]:

$$n_{TiO_2}(\lambda) = \left(4.99 + \frac{1}{96.6\lambda^{1.1}} + \frac{1}{4.60\lambda^{1.95}}\right)^{1/2} \tag{4}$$

where $\lambda$ is the wavelength [in micrometers, and $\varepsilon_{TiO_2}(\lambda) = n_{TiO_2}^2(\lambda)$].
To determine the dielectric function of the layer made of ITO nanoparticles, TiO2 nanoparticles and air, ITO:TiO2:air (we call it $\varepsilon_{eff1,\omega}$), we use a generalized Rayleigh mixing formula for a three-phase mixture [19]:

$$\frac{\varepsilon_{eff1,\omega} - \varepsilon_{Air}}{\varepsilon_{eff1,\omega} + 2\varepsilon_{Air}} = (f_{ITO} + f_{TiO_2}) \frac{f_{ITO}(\varepsilon_{ITO,\omega} - \varepsilon_{Air}) + f_{TiO_2}t_{12,\omega}(\varepsilon_{TiO_2,\omega} - \varepsilon_{Air})}{f_{ITO}(\varepsilon_{ITO,\omega} + 2\varepsilon_{Air}) + f_{TiO_2}t_{12,\omega}(\varepsilon_{TiO_2,\omega} + 2\varepsilon_{Air})} \tag{5}$$

with

$$t_{12,\omega} = \frac{3\varepsilon_{ITO,\omega}}{\varepsilon_{TiO_2,\omega} + 2\varepsilon_{ITO,\omega}} \tag{6}$$

The frequency dependent complex refractive index of the ITO:TiO$_2$:air layer is determined considering that $n_{eff1,\omega} = \sqrt{\varepsilon_{eff1,\omega}}$.

*SiO$_2$:air layer refractive index*: The wavelength dependent refractive index of SiO$_2$ can be described by the following Sellmeier equation [20]:

$$n_{SiO_2}^2(\lambda) - 1 = \frac{0.6961663\lambda^2}{\lambda^2 - 0.0684043^2} + \frac{0.4079426\lambda^2}{\lambda^2 - 0.1162414^2} + \frac{0.8974794\lambda^2}{\lambda^2 - 9.896161^2} \tag{7}$$

where $\lambda$ is the wavelength in micrometers. To determine the effective dielectric function of the SiO$_2$:air layer (we call it $\varepsilon_{eff2,\omega}$) we use the Maxwell Garnett effective medium approximation [17,21]:

$$\varepsilon_{eff2,\omega} = \varepsilon_{Air} \frac{2(1-f_{SiO_2})\varepsilon_{Air} + (1+2f_{SiO_2})\varepsilon_{SiO_2,\omega}}{2(2+f_{SiO_2})\varepsilon_{Air} + (1-f_{SiO_2})\varepsilon_{SiO_2,\omega}} \tag{8}$$

$f_{SiO_2}$ is the filling factor of the SiO$_2$ nanoparticles. In this study we choose $f_{SiO_2} = 0.7$. The photon energy dependent complex refractive index of the SiO$_2$:air layer determined considering that $n_{eff2,\omega} = \sqrt{\varepsilon_{eff2,\omega}}$.

*Transmission of the multilayer photonic crystal*: We use the two refractive indexes $n_{eff1,\omega}$ and $n_{eff2,\omega}$ for the ITO:TiO$_2$:air layers and for the SiO$_2$:air layers, respectively, to study the light transmission through the photonic structure by employ the transfer matrix method [8,22,23]. For a transverse electric (TE) wave the transfer matrix for the *k*th layer is given by

$$M_k = \begin{bmatrix} \cos\left(\frac{2\pi}{\lambda} n_k d_k\right) & -\frac{i}{n_k}\sin\left(\frac{2\pi}{\lambda} n_k d_k\right) \\ -i n_k \sin\left(\frac{2\pi}{\lambda} n_k d_k\right) & \cos\left(\frac{2\pi}{\lambda} n_k d_k\right) \end{bmatrix} \tag{9}$$

with $n_k$ the refractive index and $d_k$ the thickness of the layer. In this study the thickness of the ITO:TiO$_2$:Air layers is 185.63 nm, while the thickness of the SiO$_2$:Air layers is 242.19 nm. The product $M = M_1 \cdot M_2 \cdot \ldots \cdot M_k \cdot \ldots \cdot M_s = \begin{bmatrix} m_{11} & m_{12} \\ m_{21} & m_{22} \end{bmatrix}$ gives the matrix of the multilayer (of *s* layers). The transmission coefficient is

$$t = \frac{2n_s}{(m_{11} + m_{12}n_0)n_s + (m_{21} + m_{22}n_0)} \tag{10}$$

with $n_s$ the refractive index of the substrate (in this study $n_s = 1.46$) and $n_0$ the refractive index of air. Thus, the light transmission of the multilayer photonic crystal is

$$T = \frac{n_0}{n_s}|t|^2 \tag{11}$$

In this work the number of bilayers selected is 10, which means a 20 layers and a total thickness of about 4278 nm.

**Results and Discussion**

In Figure 1a we show a sketch of an ITO:TiO$_2$:air layer. The blue dots indicate the ITO nanoparticles, while the grey dots indicate the TiO$_2$ nanoparticles. The white interstitial volume depicts the air voids between the nanoparticles. To determine the effective refractive index of the layer we use a generalized Rayleigh equation for a three-phase mixture (Eq. 5).

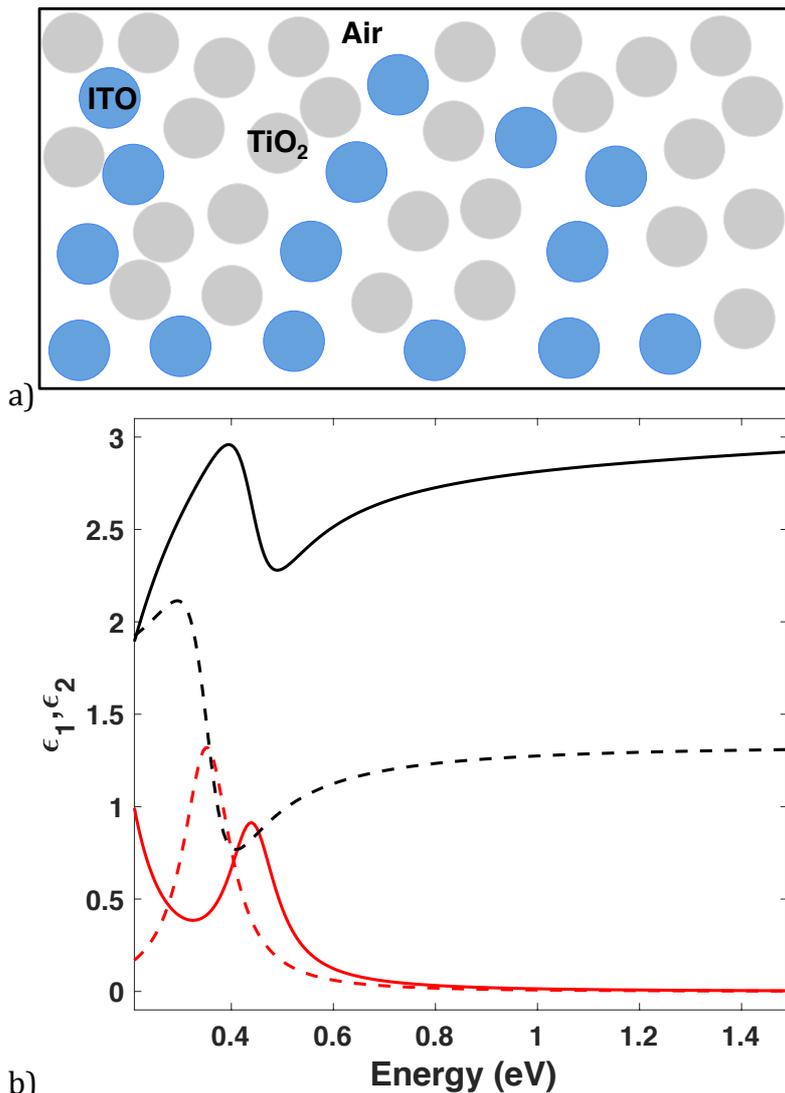

**Figure 1.** (a) Sketch of a ITO:TiO$_2$:air layer, with ITO nanoparticles (blue dots), TiO$_2$ nanoparticles (grey dots) and air (white interstitial volume); (b) Real part (solid black curve) and imaginary part (solid red curve) of the dielectric function of a ITO:TiO$_2$:air (20:50:30 ratio) layer and real part (dashed black curve) and imaginary part (dashed red curve) of a ITO:air (20:80 ratio) layer.

In Figure 1b the dashed curve represent the real (black) and imaginary (red) part of the dielectric function of an ITO:air two-phase mixture, which is a dispersion of ITO nanoparticles, with a filling factor of 0.2, in air. This two-phase mixture refractive index is determined with the Maxwell-Garnett effective medium theory [17,21]. Instead, the solid curves in Figure 1b represent the real (black) and imaginary (red) part of the effective dielectric function of an ITO:TiO$_2$:air mixture. It is evident that the ITO resonance shows a splitting and one resonance shifts towards higher energies.

In Figure 2 we show the difference between a photonic crystal with layers of silica nanoparticles alternated with layers of titania nanoparticles (black curve), and a photonic crystal with layers of silica nanoparticles alternated with a layer of ITO nanoparticles (red curve).

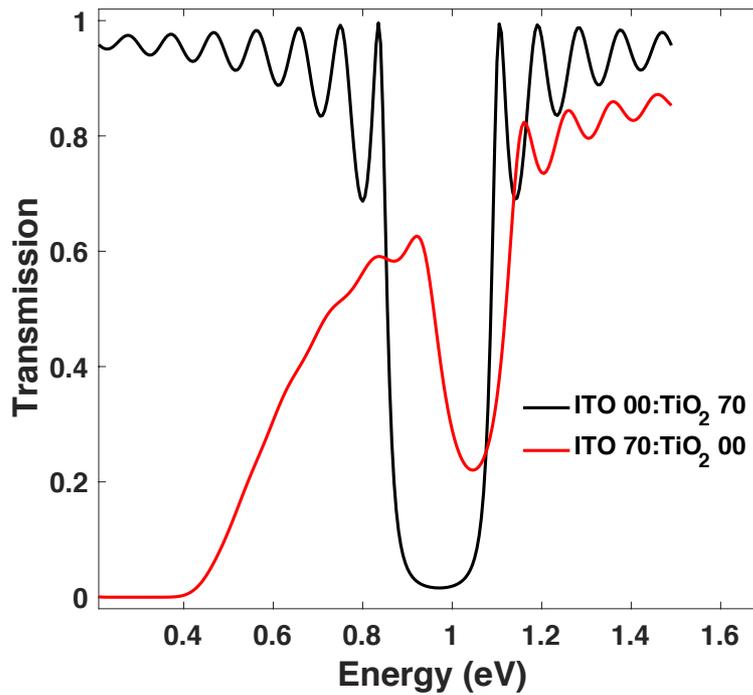

**Figure 2.** Light transmission spectrum of a SiO$_2$ nanoparticle / TiO$_2$ nanoparticle photonic crystal (black curve) both with a filling factor of 0.7, and light transmission spectrum of a SiO$_2$ nanoparticle / ITO nanoparticle photonic crystal (red curve) both with a filling factor of 0.7.

The silica/ITO photonic crystal shows a strong transmission valley at lower energy due to the plasmon resonance of ITO. Moreover, the silica/ITO photonic crystal shows a less intense photonic band gap with respect to the one of the silica/titania photonic crystal because of the smaller refractive index contrast. We can now relate to the two spectra in Figure 2 as the two extremes of a three-phase mixture of ITO nanoparticles, TiO$_2$ nanoparticles and air in which we keep constant the volume of air and we change the ITO:TiO$_2$ ratio. We show this ratio variation in Figure 3, where the increasing content of ITO relates to an increase of the transmission valley intensity due to the plasmon resonance. Noteworthy, the plasmon resonance is blue shifted with respect to the one of the ITO:air layer.

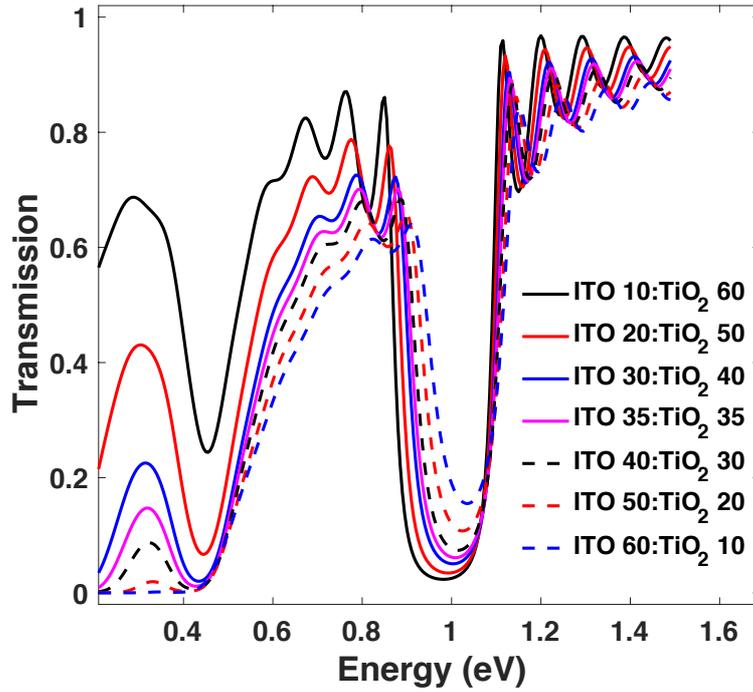

**Figure 3.** Light transmission spectra of ITO:TiO$_2$:air / SiO$_2$:air multilayer photonic crystals with different ITO:TiO$_2$ ratios, keeping constant the sum of $f_{ITO} + f_{TiO_2} = 0.7$ .

In Figure 4 we show the transmission value in correspondence of the plasmon resonance and the photonic band gap of the structure as a function of the ITO content, highlighting a cross over between $f_{ITO} = 0.2$ and $f_{ITO} = 0.3$ .

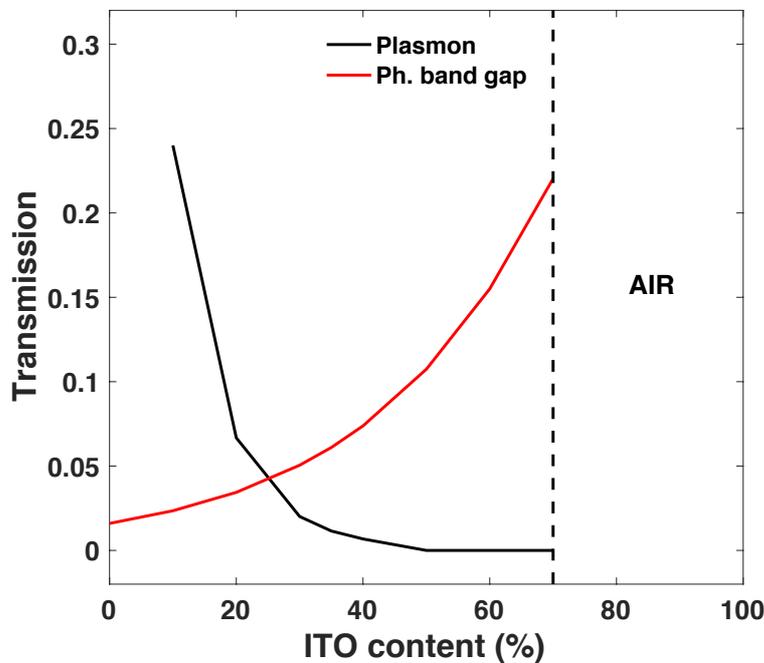

Figure 4. Transmission minima of the plasmon resonance and of the photonic band gap as a function of the ITO nanoparticles, keeping constant the sum of $f_{ITO} + f_{TiO_2} = 0.7$ .

By changing the number of carriers in ITO we can modulate the light transmission of the photonic crystal, since the change in the number of carriers leads to a change of the plasma frequency and thus the dielectric function of ITO. This results in a simultaneous change of the plasmon resonance and of the photonic band gap.

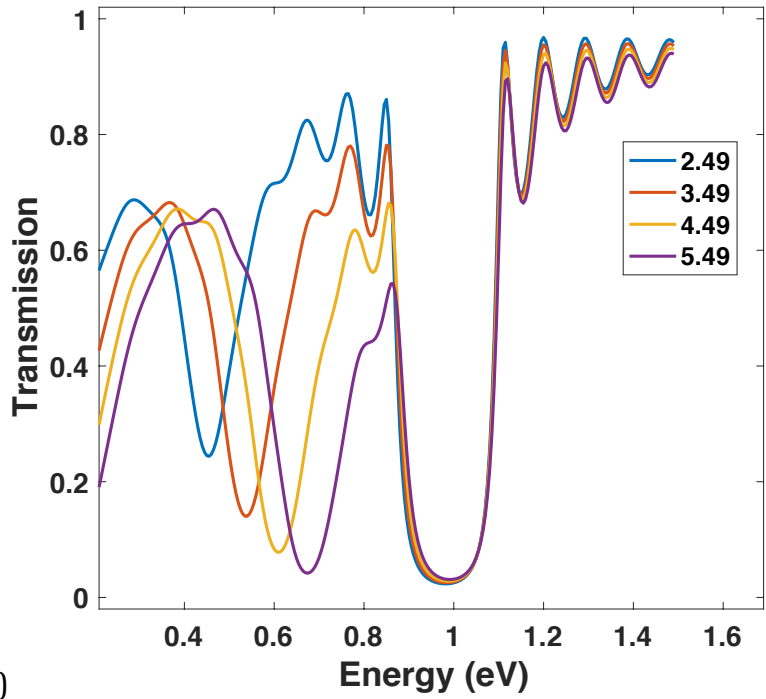

a)

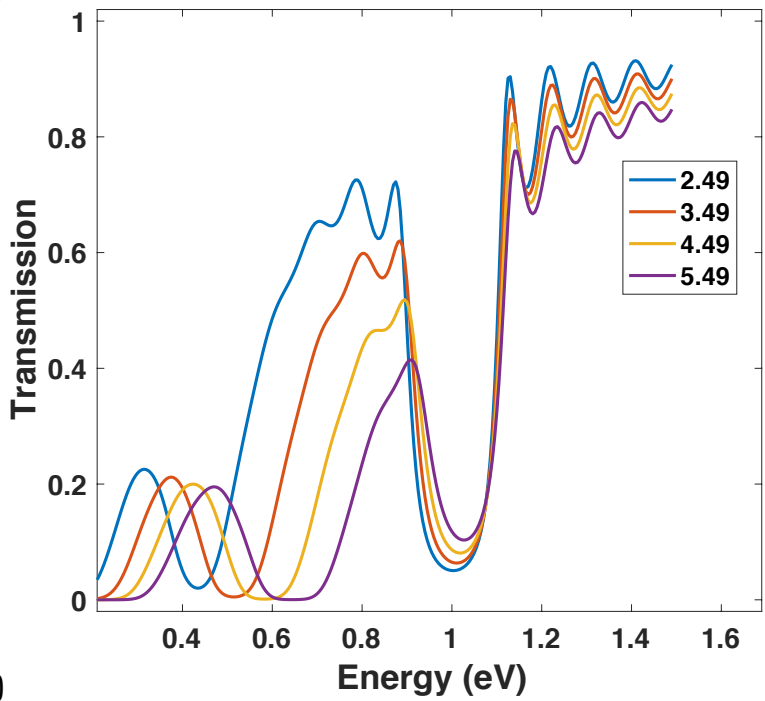

b)

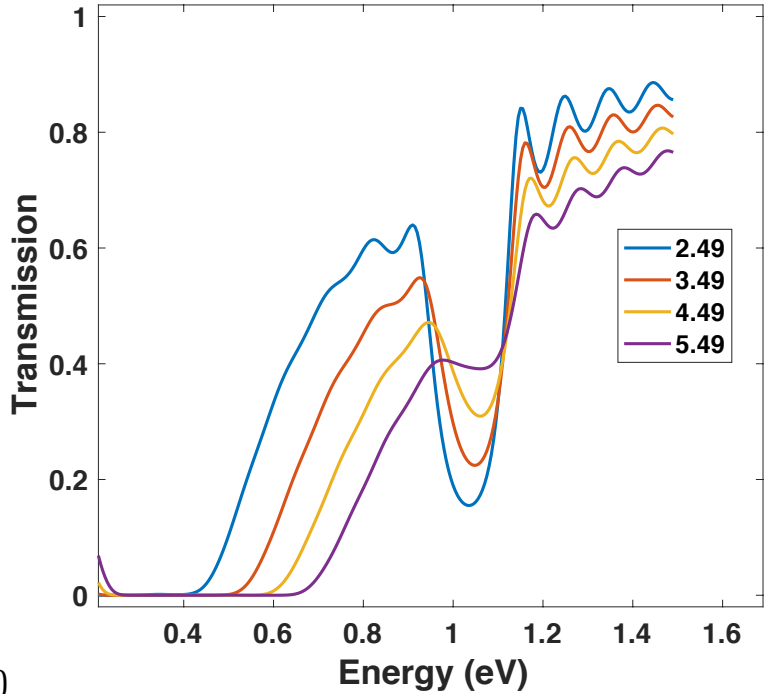

c)

**Figure 5.** Light transmission spectra of the photonic crystal as a function of the doping level ( $2.49\times10^{26}\ charges/m^3$ , $3.49\times10^{26}\ charges/m^3$ , $4.49\times10^{26}\ charges/m^3$ , $5.49\times10^{26}\ charges/m^3$) for the ITO:TiO$_2$ ratio (a) 10:60, (b) 30:40, (c) 60:10.

In Figure 5 we show the change to the modulation of the number of carriers in ITO for photonic crystals where the ratio in the ITO:TiO$_2$:air layer is 10:60:30 (Figure 5a), 30:40:30 (Figure 5b), 60:10:30 (Figure 5c). We change the number of carriers from 2.49 to $5.49\times10^{26}\ charges/m^3$ . This change in the number of carriers can be for example induced via electrochemical [16] or optical [24] way.

If we use the photonic crystal where the ITO:TiO$_2$:air layer has a ratio 60:10:30 as a filter for the Sun irradiation (direct and circumsolar [25]). We see that in the three bands in the range between 0.6 and 1.1 eV the light transmission is switched for two doping levels.

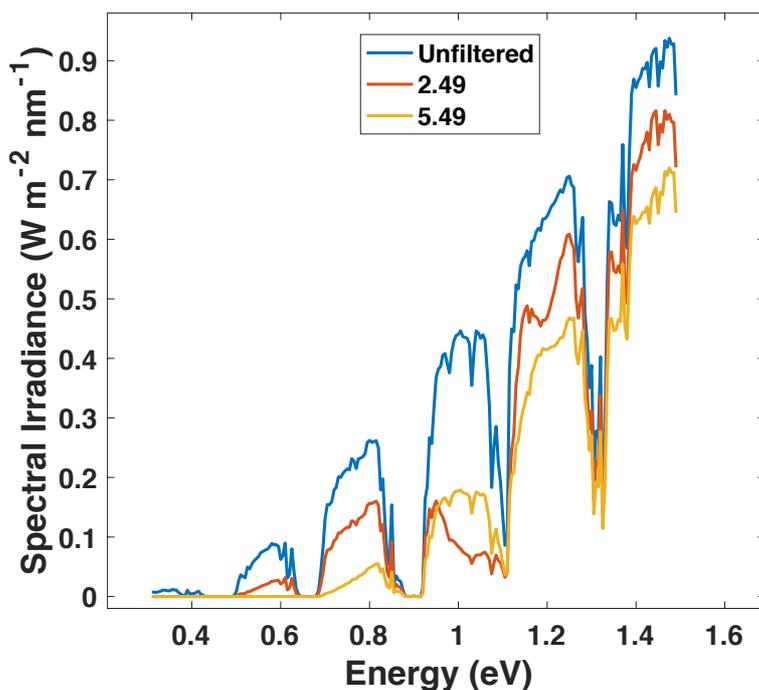

**Figure 6.** Transmission of the Sun without filter (blue curve), with an ITO:TiO$_2$ 60:10 sample characterized by a doping level of $2.49 \times 10^{26}\ charges/m^3$ (red curve) and $5.49 \times 10^{26}\ charges/m^3$ (yellow curve).

We need to consider that a layer in which ITO nanoparticles and TiO$_2$ nanoparticles are in close contact can be seen as a bulk heterojunction, with a large contact surface between the two materials, in which plasmon-induced hot electron generation can occur [26,27]. In a future work, for a more precise description of this interesting entangled multicomponent nanostructure, we should take into account this type of interaction.

**Conclusion**

In this work we show the light transmission properties of a one-dimensional multilayer photonic crystal made by alternating silica nanoparticle layers and ITO nanoparticle:TiO$_2$ nanoparticle layers. We have calculated the effective refractive index of the silica layer considering it as a two-phase mixture and the one of the ITO:titania layer considering it as a three-phase mixture. We use the calculated effective refractive indexes in the transfer matrix method to determine the light transmission of the multilayer structure. The ITO content in the ITO:TiO$_2$:air mixture affects the intensity of the plasmon resonance and the photonic bad gap. Future works can be focussed on the interaction, in terms of plasmon-induced hot electron extraction, in indium tin oxide – TiO$_2$ nanoheterojunctions.

**Acknowledgement**
This project has received funding from the European Research Council (ERC) under the European Union's Horizon 2020 research and innovation programme (grant agreement No. [816313]).